\documentclass[a4paper]{article}

\usepackage{INTERSPEECH2020}

\usepackage{float}
\usepackage{multirow}

\title{A Lite Microphone Array Beamforming Scheme with Maximum Signal-to-Noise Ratio Filter}
\name{Lu Ma, Xin Zhao, Pei Zhao, Tengrong Su}
\address{Haier Smart Home Co., Ltd.}
\email{malu@haier.com, iamroad@163.com}

\begin{document}

\maketitle
\begin{abstract}
  Since space-domain information can be utilized, microphone array beamforming is often used to enhance the quality of the speech by suppressing directional disturbance. However, with the increasing number of microphone, the complexity would be increased. In this paper, a concise beamforming scheme using Maximum Signal-to-Noise Ratio (SNR) filter is proposed to reduce the beamforming complexity. The maximum SNR filter is implemented by using the estimated direction-of-arrival (DOA) of the speech source localization (SSL) and the solving method of independent vector analysis (IVA). Our experiments show that when compared with other widely-used algorithms, the proposed algorithm obtain higher gain of signal-to-interference and noise ratio (SINR).
\end{abstract}

\section{Introduction}
Beamforming of microphone array aims to enhance the quality of a speech source. It is important for voice capture \cite{caption} in many human-computer interaction applications, such as human-robot interaction, camera steering and intelligent monitoring.

There are many beamforming algorithms, such as delay-and-sum (DS) beamforming \cite{DS, DS1}, linear constraint minimal variance (LCMV) \cite{LCMV}, minimum variance distortionless response (MVDR) \cite{MVDR, MVDR1}, and blind source separation. Generally, the spacing between microphones of the smart devices is relatively small, resulting in few difference of sampling points between microphones. This would lead to low precision of beamforming, restricting the application of DS beamforming.

The MVDR technique is perhaps the most widely used adaptive beamformer. The basic underlying idea is to choose the coefficients of the filter that minimize the output power with the constraint that the desired signal is not affected. The advantage of the MVDR over a DS beamformer is that this adaptive beamformer can adapt itself to the noise environment for maximum noise reduction. Unfortunately, the performance of MVDR would degrade when the number of microphones decreased or the spacing between the microphones diminished.

However, blind source separation (BSS) is a powerful technique to find independent components from mixtures without mixing information, which has been widely used for beamforming \cite{BSS}. One of the popular algorithms is the independent vector analysis (IVA) \cite{IVA}, where a multivariate function is adopt as nonlinear score function, which can use the data from all the frequency bins compared with the independent component analysis (ICA) \cite{ICA} with a univariate function where only the data in each frequency bin are used to update the un-mixing matrix. Whereas the iteration process for calculating the weight coefficients would increase the complexity of the IVA algorithm, especially when the number of microphone increases.

Moreover, with the increasing number of the microphone, the computation complexity of all the above methods except the DS method will increase due to large matrix computation. In this context, long computation delay arises or high performance hardware is required.

In this paper, a lite beamforming scheme based on maximum signal-to-noise ratio (SNR) criterion \cite{MAXSNR} is proposed to decrease the complexity by employing the direction-of-arrival (DOA) information of speech source localization (SSL) \cite{SSL} and we will show that, in some cases, the the maximum SNR filter can be equivalent to IVA from the perspective of mathematics. Moreover, a straightforward beamforming scheme by selecting the optimum pair of microphones based on the DOA result of SSL is proposed to further reduce the computation complexity for circular arrays.

\section{Maximum Signal-to-Noise Ratio Filter}
\subsection{Algorithm Principle}
The received signal ${x_{nk}}\left( {n = 0,1, \ldots ,N - 1} \right)$ of array element at time $k$ can be expressed as a vector by
\begin{equation}
\begin{array}{*{20}{l}}
{{x_k} = {{\left[ {{x_{0k}}\begin{array}{*{20}{c}}
{}
\end{array}{x_{1k}}\begin{array}{*{20}{c}}
 \cdots
\end{array}{x_{\left( {N - 1} \right)k}}} \right]}^{\rm{T}}}}\\
{\begin{array}{*{20}{c}}
{}
\end{array} = {a_k}\left[ {1\begin{array}{*{20}{c}}
{}
\end{array}{e^{ - j\frac{{2\pi d}}{\lambda }\sin \theta }}\begin{array}{*{20}{c}}
 \cdots
\end{array}{e^{ - j\left( {N - 1} \right)\frac{{2\pi d}}{\lambda }\sin \theta }}} \right]}\\
{\begin{array}{*{20}{c}}
{}&{}&{}
\end{array}\begin{array}{*{20}{c}}
{}&{}&{}
\end{array}\begin{array}{*{20}{c}}
{}&{}
\end{array} + {{\left[ {{v_{0k}}\begin{array}{*{20}{c}}
{}
\end{array}{v_{1k}}\begin{array}{*{20}{c}}
 \cdots
\end{array}{v_{\left( {N - 1} \right)k}}} \right]}^{\rm{T}}}}\\
{\begin{array}{*{20}{c}}
{}
\end{array} = {a_k}s\left( \theta  \right) + {v_k}}
\end{array}
\label{eq1}
\end{equation}
where, ${a_k}$ is the sample of complex envelope ${a_{nk}}$ of target signal received by each array element, $s\left( \theta  \right)$ is the direction vector of the target signal, which contains the direction information of the target signal received by the array element and is independent of time, and ${a_k}s\left( \theta  \right)$ is the target signal vector, ${v_k}$ is the zero mean stationary additive external interference plus internal noise vector. The autocorrelation matrix of the signal vector is
\begin{equation}
\begin{array}{l}
{R_x} = E\left[ {{x_k}x_k^{\rm{H}}} \right] = E\left\{ {\left[ {{a_k}s\left( \theta  \right) + {v_k}} \right]{{\left[ {{a_k}s\left( \theta  \right) + {v_k}} \right]}^{\rm{H}}}} \right\}\\
\begin{array}{*{20}{c}}
{}
\end{array} = E\left[ {{a_k}s\left( \theta  \right){s^{\rm{H}}}\left( \theta  \right)a_k^{\rm{H}}} \right] + E\left[ {{v_k}v_k^{\rm{H}}} \right]\\
\begin{array}{*{20}{c}}
{}
\end{array} = \sigma _s^2s\left( \theta  \right){s^{\rm{H}}}\left( \theta  \right) + {R_v} = {R_s} + {R_v}
\end{array}
\label{eq2}
\end{equation}
where, ${x_k^{\rm{H}}}$ is the conjugate transpose of the received signal vector; ${{s^{\rm{H}}}\left( \theta  \right)}$ is the conjugate transpose of the steering vector of the target signal. The autocorrelation matrix of the target signal vector is ${R_s} = \sigma _s^2s\left( \theta  \right){s^{\rm{H}}}\left( \theta  \right)$ and $\sigma _s^2 = E\left[ {{a_k}a_k^{\rm{H}}} \right] = E\left[ {a{a^{\rm{H}}}} \right]$ , ${v_k}$ is the conjugate transpose of the interference plus noise vector, ${R_v} = E\left[ {{v_k}v_k^{\rm{H}}} \right] = E\left[ {v{v^{\rm{H}}}} \right]$ is the covariance matrix of the interference plus noise vector, the target signal vector ${a_k}s\left( \theta  \right)$  is independent of the noise vector $v$. Denoting the weighted by
\begin{equation}
{w_k} = {\left[ {{w_{0k}}\begin{array}{*{20}{c}}
{}
\end{array}{w_{1k}}\begin{array}{*{20}{c}}
 \cdots
\end{array}{w_{\left( {N - 1} \right)k}}} \right]^{\rm{T}}}
\label{eq3}
\end{equation}

Then the sum of weighted array signals is the beamforming output signal, namely
\begin{equation}
{y_k}\left( \theta  \right) = \sum\limits_{n = 0}^{N - 1} {w_{nk}^*{x_{nk}}}  = w_k^{\rm{H}}{x_k} = w_k^{\rm{H}}\left[ {{a_k}s\left( \theta  \right) + {v_k}} \right]
\label{eq4}
\end{equation}

Thus, after beamforming, the power ratio between target signal and interference plus noise can be obtained as follows
\begin{equation}
{\rm{SNR}} = \frac{{{P_s}}}{{{P_v}}} = \frac{{w_k^{\rm{H}}{R_s}{w_k}}}{{w_k^{\rm{H}}{R_v}{w_k}}}
\label{eq7}
\end{equation}
where ${P_s} = w_k^{\rm{H}}{R_s}{w_k}$ is the average output power of the target signal after beamforming, and ${P_v} = w_k^{\rm{H}}{R_v}{w_k}$ is the average output power of the interference plus noise vector.

With maximum SNR criterion, the largest weighted vector is the optimal weighted vector ${w_{{\rm{opt}}}}$ that make the SNR maximal. It is the eigenvector corresponding to the maximum generalized eigenvalue of the autocorrelation matrix of ${x_k}$ with respect to $\left( {{R_s},{R_v}} \right)$, and can be expressed as
\begin{equation}
{R_s}{w_{k,{\rm{opt}}}} = {\lambda _{\max }}{R_v}{w_{k{\rm{,opt}}}}
\label{eq8}
\end{equation}

According to Eq. (\ref{eq8}), in order to obtain optimal weighting vector, the autocorrelation of source and the cross-correlation of interference plus noise should be known in advance. However, the signal received by microphone is a mixture of source signal and interference plus noise, it is difficult to separate these two kinds of signals without any prior information. Here, we will deduce that this problem can fortunately be solved from the perspective of Independent Vector Analysis (IVA).

\subsection{Independent Vector Analysis}
One variant of IVA is called the auxiliary function approach, named AuxIVA which adopts the auxiliary function technique to avoid the step size tuning \cite{IVA2}. In the auxiliary function technique, an auxiliary function is designed for optimization. During the learning process, the auxiliary function is minimized in terms of auxiliary variables. The auxiliary function technique can guarantee monotonic decrease of the cost function, and therefore provides effective iterative update rules. The optimal beamforming weights is to solve the following simultaneous vector equations \cite{IVA2}.
\begin{equation}
w_l^h{V_k}{w_k} = {\delta _{lk}}\begin{array}{*{20}{c}}
{}
\end{array}\left( {1 \le k \le K,1 \le l \le K} \right)
\label{eq9}
\end{equation}
where ${w_l}$ and ${w_k}$ are weighting coefficients, ${V_k}$ is auxiliary variable,   is the number of source. Unfortunately, there are no closed-form solutions for eq. \ref{eq9} considering updating all of ${w_k}$ simultaneously, except when $K = 2$.

When $K = 2$, Eq. (\ref{eq9}) indicates that both of ${V_1}{w_1}$ and ${V_2}{w_1}$ are orthogonal to ${w_2}$. Because the direction orthogonal to ${w_2}$ is uniquely determined in the two dimensional space, ${V_1}{w_1}$ and ${V_2}{w_1}$ have to be parallel such as
\begin{equation}
{V_1}{w_1} = \gamma {V_2}{w_1}
\label{eq10}
\end{equation}
where $\gamma $ is a constant. In the same way, ${V_1}{w_1}$ and ${V_2}{w_1}$ are also parallel. Such vectors are obtained as solutions of Eq. (\ref{eq9}), which is a generalized eigenvalue problem.

By comparing Eq. (\ref{eq10}) with Eq. (\ref{eq8}), we can find that they have the same expression form the perspective of mathematics. This reminds us that the solutions of the maximum SNR filter could be solved with the AuxIVA in this circumstance.

\subsection{Maximum SNR Filter Realization}
By associating Eq. (\ref{eq10}) with Eq. (\ref{eq8}), we can deduce that the expressions for autocorrelation matrix of the source ${R_s}$ and the interference ${R_v}$ as
\begin{equation}
{R_s} = {V_1},{R_v} = {V_2}
\label{eq11}
\end{equation}
It can be found that the key point is to calculate the auxiliary variable ${V_k}$, which can be expressed by
\begin{equation}
{V_i}\left( k \right) = E\left[ {\frac{{{{G'}_R}\left( {{r_i}} \right)}}{{{r_i}}}{\bf{x}}\left( k \right){\bf{x}}{{\left( k \right)}^\dag }} \right]
\label{eq12}
\end{equation}
where ${\bf{x}}\left( k \right)$ is the receiving vector of the microphone array, ${G_R}\left( {{r_i}} \right)$ is the contrast function of AuxIVA, and ${r_i} = {\left\| {{{\hat s}_i}} \right\|_2}$. When employing multivariate generalized Gaussian distribution as the source prior and considering energy correlation between different frequency bins, meanwhile making it more robust to outliers, the following contrast function can be deduced,
\begin{equation}
G\left( {{{\hat s}_i}} \right) = {G_R}\left( {{r_i}} \right) = r_i^{\frac{2}{3}},\begin{array}{*{20}{c}}
{}
\end{array}{r_i} = \sqrt {\sum\limits_{k = 1}^K {{{\left| {{\bf{w}}_i^\dag \left( k \right){\bf{x}}\left( k \right)} \right|}^2}} }
\label{eq13}
\end{equation}
By substituting the formula ${r_i}$ into $\frac{{{{G'}_R}\left( {{r_i}} \right)}}{{{r_i}}}$, we have
\begin{equation}
\frac{{{{G'}_R}\left( {{r_i}} \right)}}{{{r_i}}} = \frac{2}{{3r_i^{\frac{4}{3}}}} = \frac{2}{{3\sqrt[3]{{\left( {\sum\limits_{k = 1}^K {{{\left| {{{\hat s}_i}\left( k \right)} \right|}^2}} } \right)}}}}
\label{eq14}
\end{equation}

In order to realize the expectation $E\left[  \cdot  \right]$, smoothing between adjacent frames is performed for ${V_i}\left( k \right)$, i.e.,
\begin{equation}
{V_i}\left( n \right) = \left( {1 - \beta } \right){V_i}\left( n \right) + \beta {V_i}\left( {n - 1} \right)
\label{eq16}
\end{equation}
where ${V_i}\left( n \right)$ and ${V_i}\left( {n - 1} \right)$ are the auxiliary variables of the n-th frame and the (n-1)-th frame, $\beta $ is the smoothing coefficient.

\begin{figure}[H]
\centering
\includegraphics[scale=0.73]{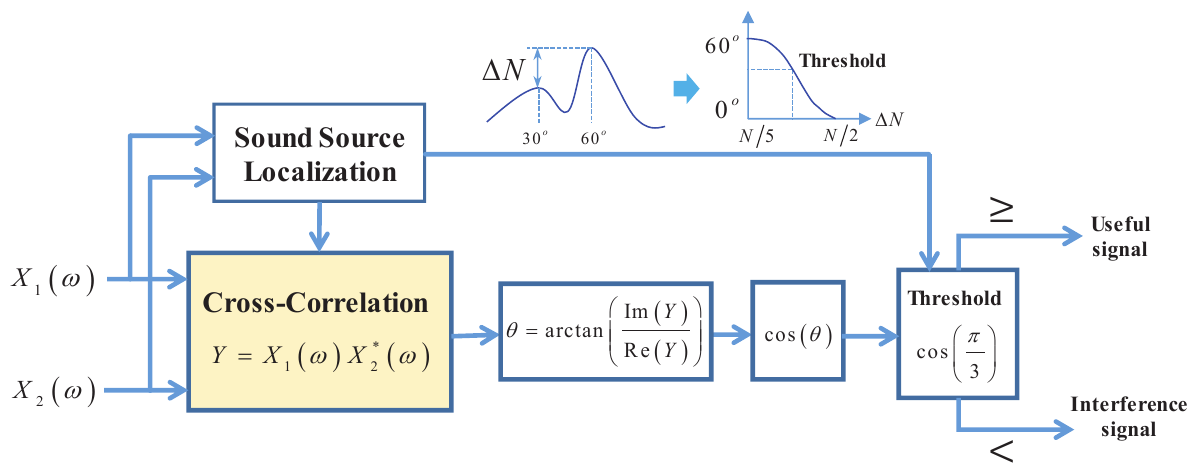}
\caption{Energy calculation for target and interference signals}
\label{fig_energy}
\end{figure}

Therefore, as long as the contrast function $\frac{{{{G'}_R}\left( {{r_i}} \right)}}{{{r_i}}}$ is obtained, would the weight coefficients be calculated. The key point for calculating the contrast function for a source is to obtain the energy of that source, i.e., $\sum\limits_{k = 1}^K {{{\left| {{{\hat s}_i}\left( k \right)} \right|}^2}}$. In the standard AuxIVA algorithm, this energy is calculated using equation ${\hat s_i}\left( k \right) = {{\bf{w}}_i}\left( k \right){\bf{x}}\left( k \right)$ by randomly selecting an initialized weight ${{\bf{w}}_i}\left( k \right)$. Then the weight ${{\bf{w}}_i}\left( k \right)$ is iteratively updated. According to Ref. \cite{mask, sepration}, the energy ratio between different sources is strongly correlated with the sound arrival time difference between the two channels. Inspired by this idea, here, we propose to speeding up the iteration by employing the DOA result of the SSL. Specifically, for a pair of microphones, the energy of the target and the interference signals can be calculated as follows and depicted in Fig. \ref{fig_energy}.

\begin{itemize}
\item Convert the two microphone reception signals to the frequency domain, denoted by ${X_1}\left( \omega  \right)$  and ${X_2}\left( \omega  \right)$ ;
\item Sound source localization is performed to get the steering vector which is used for phase compensation between these two microphones and expressed as
\begin{equation}
{s_k} = \left[ {\begin{array}{*{20}{c}}
1&{{e^{ - j\Delta \varphi }}}
\end{array}} \right],\begin{array}{*{20}{c}}
{}&{\Delta \varphi  = \frac{{2\pi {f_k}d\sin \theta }}{c}}
\end{array}
\label{eq17}
\end{equation}

where ${s_k}$ is the direct vector for these two microphones, $\Delta \varphi$ is the phase difference between the microphones, ${f_k}$ is the frequency of the $k$-th bin, $\theta$ is the DOA of the target source, $c$ is the speed of sound, $d$ is the spacing between the two microphones.

\item Calculate the cross-correlation between two complex signals of microphones for each frequency bin as $Y = {X_1}\left( \omega  \right)X_2^ * \left( \omega  \right)$;
\item Extract the phase difference between microphones by $\theta  = \arctan \left( {\frac{{{\mathop{\rm Im}\nolimits} \left( Y \right)}}{{{\mathop{\rm Re}\nolimits} \left( Y \right)}}} \right)$, where ${{\mathop{\rm Im}\nolimits} \left( Y \right)}$ and ${{{\mathop{\rm Re}\nolimits} \left( Y \right)}}$ are the imaginary and real parts of $Y$;
\item Calculate the cosine value of the phase difference and compare it with the threshold. The frequency bin belongs to useful signal namely the target signal when the cosine value is greater than the threshold, otherwise is interference signal. These two sets of bins can be denoted by $T$ and $I$ respectively.  Then the energy of the target and the interference signals can be calculated by
\begin{equation}
{E_T} = \sum\limits_{k = 1,k \in T}^K {{{\left| {{X_1}\left( k \right)} \right|}^2}} ,{E_I} = \sum\limits_{k = 1,k \in I}^K {{{\left| {{X_2}\left( k \right)} \right|}^2}}
\label{eq15}
\end{equation}

\end{itemize}

The threshold can be adjusted according to the environment for robustness. One adaptive method is to utilize the localization results as depicted in Fig. \ref{fig_energy}. For localization, the number of bins belongs to the corresponding DOA is accumulated, and the one with maximum number is selected as the result. In this way, the number of the second largest could also be obtained, and the gap between these two numbers is calculated. It is obvious that the stronger the interference is, the bigger the gap will be. Therefore, the threshold can be associated with the gap. This is done by constructing a mapping function between these two quantities. A feasible way depicted in Fig. \ref{fig_energy} is exponential descent function (EDF). The scope of the x-axis and that of the y-axis are from ${N \mathord{\left/
 {\vphantom {N 5}} \right.\kern-\nulldelimiterspace} 5}$ to ${N \mathord{\left/
 {\vphantom {N 5}} \right.\kern-\nulldelimiterspace} 2}$ and from $0^o$ to $60^o$, respectively.

Here the threshold can be considered as beamwidth (assumed to be ${60^o}$ in Fig. \ref{fig_energy}). If the phase difference falls within the beamwidth, it is a useful signal, others is interference. In this way, contrast variable $\frac{{{{G'}_R}\left( {{r_i}} \right)}}{{{r_i}}}$ can be obtained, so as to get the covariance matrices ${V_i}$.


\section{Lite Beamforming Scheme}
\subsection{Sound Source Localization}
\begin{figure}[H]
\centering
\includegraphics[scale=0.75]{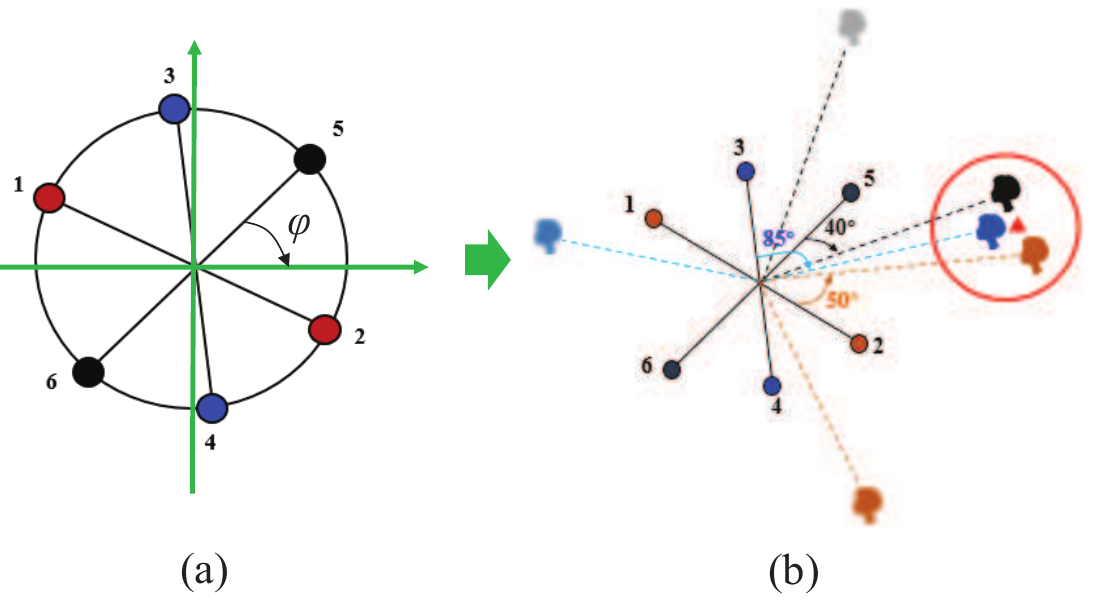}
\caption{Sound source localization for circular array}
\label{fig_circularDOA}
\end{figure}

Localization with dual-microphone can employ the method proposed in Ref. \cite{SSL}, which produces more reliable performance in realistic reverberant environments compared with other widely-used algorithms. To further employing this method to circular array, a trivial method is proposed here to improve the accuracy of localization of circular array. This can be illustrated in Fig. \ref{fig_circularDOA}. For circular array, the two microphones whose connecting line passing across the center are considered as a pair, and are used for localization with the method of Ref. \cite{SSL}. This can be performed by three steps:
\begin{enumerate}
\item Localization with dual-microphones. This is performed by the method of Ref. \cite{SSL}. Due to the geometric symmetry constructed by dual-microphones, a mirror source is also obtained apart from the real one.
\item Angle calibration. After localization with each pair of microphones, angle calibration is performed to convert the coordinate system to the global one. This is done by rotating the localization result by an angle of $\varphi$ as depicted in Fig. \ref{fig_circularDOA}(a). For each pair of microphones, the angle with respect to the global coordinate system can be calculated by $
\varphi  = {\tan ^{ - 1}}\left( {\frac{{{y_1} - {y_0}}}{{{x_1} - {x_0}}}} \right)
$.
\item Mirror elimination. As long as the localization angles with respect to the global coordinate system are obtained, the mirrors can be eliminated by clustering these angles and the real angle is the one with maximal members. This can be depicted by Fig. \ref{fig_circularDOA}(b). Since the dual-microphones localization is realized by counting the results of each frequency bins, the localization angle of the circular array can also be obtained by counting the number of unambiguous angles of all frequency bins of these three pairs of microphones. The real localization result is the center of the angle scope with maximum number of angles. Another way is to consider the center of the cluster comprised by all the unambiguous angles of each frequency bin of these three pairs of microphones.
\end{enumerate}

\subsection{Beamforming Scheme}
For simplicity and low complexity, only one pair of microphones could be selected for beamforming. This can be done by calculating the angle difference between the DOA obtained by the circular array and the normal direction perpendicular to the connecting line of each microphone pair. The one with the minimal angle difference is selected as the microphone pair for beamforming based on maximum SNR method. For example, as depecited in Fig. \ref{fig_circularDOA}(b), since the angle $85^o$ of the pair denoted by $\# 3$ and $\#4$ is more closer to the normal of its connecting line compared with other pairs, this pair would be selected for beamforming as illustrated in Fig. \ref{fig_circularDOA}(b). In this time, the phase of this pair should be aligned, i.e., multiplying by steering vector. This is done by compensating the phase between these two microphones with DOA as ${s_k} = \frac{1}{2}{\left[ {{e^{j\frac{{\pi d}}{\lambda }\sin \theta }}\begin{array}{*{20}{c}}
{}
\end{array}{e^{ - j\frac{{\pi d}}{\lambda }\sin \theta }}} \right]^{\rm{T}}}
\label{eq20}
$

In this way, the phase of the two microphones are aligned to the center of the array, and no phase offset would arise whichever pair is selected. After compensition, the aforementioned beamforming with maximum SNR can be performed.

\section{Performance Evaluation}
\subsection{Dual array}
\begin{figure}[H]
\centering
\includegraphics[scale=1.0]{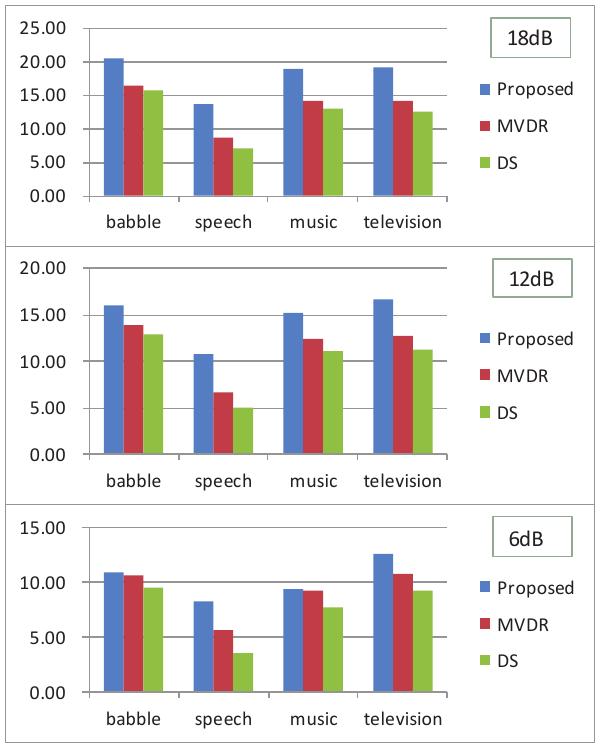}
\caption{SINR gain obtained at different input SINR}
\label{fig_lineargain}
\end{figure}

Assuming the microphone spacing is 85 mm. The signal-to-interference and noise ratio (SINR) gain obtained by beamforming under different interference types and different input SNR is shown in Fig. \ref{fig_lineargain}. As can be seen from the figures, for different types of interference, the proposed method has higher SINR gain compared with other conventional method. This gain would increase with the increasing SINR of input. This is because that with the increasing SINR of input, the accuracy of source localization would increase and so is to the energy calculation as expressed in Eq. (\ref{eq15}).

\subsection{Circular array}
\begin{figure}
\centering
\includegraphics[scale=1.1]{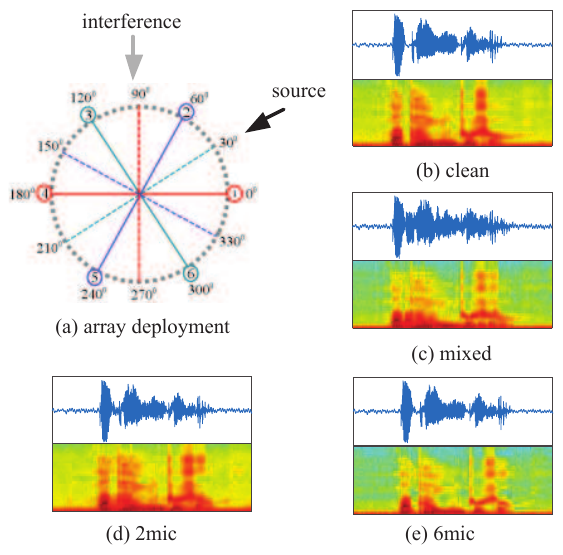}
\caption{Array deployments for simulation}
\label{fig_compare}
\end{figure}

Since the beamforming of circular array is realized by selecting the best pair of microphones whose normal is closest to the sound source direction, it shall have higher performance than that of the two microphones itself. This can be illustrated in Fig. \ref{fig_compare}, where the source and the interference are locating at $30^o$ and $90^o$ respectively and the microphone pair of $\# 1$ and $\# 4$ is used for dual-mic beamforming. As long as the DOA of circular array is calculated, the pair of $\# 3$ and $\# 6$ would be selected as the best pair for beamforming. Moreover, since the localization of circular array could be more accuracy than the dual-mic itself, the energy calculation for maximum SNR beamforming can be more accuracy than the dual-mic itself. These all results in higher SINR gain for circular array.

The SINR gain obtained at different orientations for dual-mic and circular array are listed in Table \ref{tab:compare1} for comparison where the input SINR is 6dB. Five directions, i.e., $30^o$, $60^o$, $90^o$, $120^o$, $150^o$ are considered. The averaged gain at one source direction is calculated by averaging the gain obtained at other directions. For example, when the source is locating at $30^o$ and the interference (or the noise) is locating at $60^o$, a SINR gain can be calculated, in the same way, when considering other directional interferences, the corresponding SINR gain could also be calculated, thus by averaging the four results, the averaged SINR for source locating at $30^o$ is obtained. It reveals that beamforming with circular would gain higher performance to suppress interference compared with that of 2-mic scheme.

\begin{table}[H]
  \caption{Comparison between 2-mic and circular array}
  \label{tab:compare1}
  \centering
\begin{tabular}{|c|c|c|c|c|}
\hline
\textbf{Source} & \textbf{Noise} & \textbf{Dual}  & \textbf{Circular}  & \textbf{Averaged}  \\
\hline
\multirow{4}*{$30^\circ$}  & $60^\circ$  & 2.14  & 11.30 & \multirow{4}{*}{8.69 vs. 13.50}    \\
    \cline{2-4}
        & $90^\circ$  & 4.02  & 13.41  & \multirow{2}{*}{} \\
    \cline{2-4}
        & $120^\circ$ & 15.46 & 14.46  & \multirow{2}{*}{}  \\
    \cline{2-4}
        & $150^\circ$ & 15.27 & 14.83  & \multirow{2}{*}{}  \\
    \hline
    \multirow{4}*{$60^\circ$}  & $30^\circ$  & 9.24  & 9.10  & \multirow{4}{*}{11.37 vs. 13.15}  \\
    \cline{2-4}
      & $90^\circ$  & 2.01  & 10.87 & \multirow{2}{*}{}  \\
    \cline{2-4}
      & $120^\circ$ & 17.62 & 17.03 & \multirow{2}{*}{}  \\
    \cline{2-4}
      & $150^\circ$ & 16.59 & 15.59 & \multirow{2}{*}{}   \\
    \hline
    \multirow{4}*{$90^\circ$}  & $30^\circ$  & 14.62 & 14.86 & \multirow{4}{*}{8.94 vs. 13.06}  \\
    \cline{2-4}
      & $60^\circ$  & 2.42  & 11.12 & \multirow{2}{*}{}  \\
    \cline{2-4}
      & $120^\circ$ & 3.06  & 11.29 & \multirow{2}{*}{}  \\
    \cline{2-4}
      & $150^\circ$ & 15.64 & 14.96 & \multirow{2}{*}{}  \\
    \hline
    \multirow{4}*{$120^\circ$}  & $30^\circ$  & 11.39 & 8.07  & \multirow{4}{*}{6.13 vs. 9.61}  \\
    \cline{2-4}
      & $60^\circ$  & 2.59 & 9.95  & \multirow{2}{*}{}  \\
    \cline{2-4}
      & $90^\circ$ & 0.63 & 8.03  & \multirow{2}{*}{}   \\
    \cline{2-4}
      & $150^\circ$ & 9.90 & 12.37  & \multirow{2}{*}{} \\
    \hline
    \multirow{4}*{$150^\circ$}  & $30^\circ$  & 10.18 & 15.11 & \multirow{4}*{6.97 vs. 14.08}  \\
    \cline{2-4}
      & $60^\circ$  & 9.70  & 14.82  & \multirow{2}{*}{}  \\
    \cline{2-4}
      & $90^\circ$  & 1.11  & 14.96  & \multirow{2}{*}{}  \\
    \cline{2-4}
      & $120^\circ$ & 6.91  & 11.43  & \multirow{2}{*}{}  \\
    \hline
\end{tabular}
\end{table}

\section{Conclusions}
An enhanced beamforming method based on maximum SNR is proposed to pick up speech signal efficiently in this paper. As long as the phase compensation among the microphones is performed by multiplying with steering vector, could the inter-channel phase difference between microphones be used for calculating the covariance matrix of source and  interference by employing the solving strategies of IVA with auxiliary function and multivariate generalized Gaussian distribution as the source prior. Therefore, generalized eigenvalue decomposition is utilized to compute the optimal weighting vector. Experiment results reveal that higher SINR gain can be obtained with the proposed method compared with other conventional method.

%

\bibliographystyle{IEEEtran}

\bibliography{mybib}


\end{document}